\documentclass[pre,showpacs,preprint]{revtex4}

\usepackage{amsmath}

\usepackage{graphicx}
\usepackage{dcolumn}
\usepackage{bm}

\begin{document}

\title{Breakdown of hydrodynamics in the inelastic Maxwell model of granular gases}
\author{J. Javier Brey, M.I. Garc\'{\i}a de Soria, and P. Maynar}
\affiliation{F\'{\i}sica Te\'{o}rica, Universidad de Sevilla,
Apartado de Correos 1065, E-41080, Sevilla, Spain}
\date{\today }

\begin{abstract}
Both the right and left eigenfunctions and eigenvalues of the linearized homogeneous Boltzmann equation for inelastic Maxwell molecules corresponding to the hydrodynamic modes are calculated. Also, some non-hydrodynamic
modes are identified. It is shown that below a critical value of the parameter characterizing the inelasticity,
one of the kinetic modes decays slower than one of the hydrodynamic ones. As a consequence, a closed hydrodynamic description does not exist in that regime. Some implications of this behavior on the formally computed Navier-Stokes transport coefficients are discussed.
\end{abstract}


\maketitle

\section{Introduction}
\label{s1}
Granular gases provide an appropriate context in which a number of fundamental issues
related with non-equilibrium systems can be addressed. Primary among these is the
existence of a hydrodynamic description, i.e. of closed, deterministic equations for the
hydrodynamic fields, identified as the number density $n$, the flow velocity ${\bm u}$, and the (granular) temperature $T$, in the case of one-component systems. A theoretical framework for deriving macroscopic equations from
the underlying microscopic dynamics and investigating its range of validity is kinetic theory. In the low density limit, the Boltzmann equation for smooth and inelastic hard spheres or disks has been used to derive hydrodynamic equations
for granular gases since many years ago \cite{LSJyCh84,JyR85,SyG98}. Presently, explicit expressions for the transport
coefficients appearing in the analogous of the Navier-Stokes equations are available, and they have been confirmed via direct Monte-Carlo simulations \cite{BDKyS98,ByC01}. Nevertheless, the methods used are formal and do not address
either the existence or the context of the hydrodynamic description itself, although internal mathematical
consistency is accomplished at the level of the Navier-Stokes approximation, i.e. first order in the gradients of the
hydrodynamic fluxes. For elastic hard spheres or disks, the problem has been satisfactorily
solved by analyzing the spectrum of the linearized Boltzmann operator \cite{Mc89}. On the other hand,
for the inelastic Boltzmann equation (IBE), although the hydrodynamic eigenfunctions and eigenvalues
have been identified in the limit of large wave vectors \cite{BDyR03,DyB03}, almost nothing is known  about the kinetic, non-hydrodynamic part of the spectrum. Consequently, the dominance of the hydrodynamic modes has not been established for the IBE, although it has been proven for some single-time relaxation model equations
\cite{DyB03}.

Here the validity of a hydrodynamic description for granular gases will be addressed using a simplified IBE. Specifically, the inelastic Maxwell model (IMM) will be employed \cite{BCyG00,CCyG00,EyB02,ByK02}.  This kinetic equation is obtained from the IBE for hard spheres of disks by replacing  the velocity dependent collision rate by an effective average value proportional to the thermal velocity. Although other choices have also been considered \cite{BCyG00,CCyG00}, here the effective collision rate will be assumed to be also independent of the collision angle. Moreover, attention will be restricted to the modes at asymptotically long wave lengths, i.e. to perturbations occurring also in homogeneous systems. This will suffice to establish that the hydrodynamic spectrum is not isolated from the rest at strong inelasticity, contrary to what is required for the existence of hydrodynamics.

Let $f({\bm v}_{1},t)$ be the one-particle distribution function of particles of mass $m$ with velocity ${\bm v}_{1}$ at time $t$. The homogenous Boltzmann equation for the the IMM in $d$-dimensions reads
\begin{equation}
\label{1.1}
\partial_{t} f({\bm v}_{1},t) =J_{M}[{\bm v}_{1},t|f,f],
\end{equation}
\begin{equation}
\label{1.1a}
J_{M}[{\bm v}_{1},t|f,g] \equiv \frac{d+2}{2} \frac{\nu (t)}{n \Omega_{d}}
\int d \widehat{\bm \sigma} \int d{\bm v}_{2} \left (\alpha^{-1} b_{\bm \sigma}^{-1}-1 \right) f({\bm v}_{1},t) g({\bm v}_{2},t).
\end{equation}
Here $\nu (t)$ is an effective collision frequency. Its explicit form will not be needed, being enough to know that $\nu \propto nT^{1/2}$. Moreover, $\Omega_{d} \equiv 2 \pi^{d/2}/\Gamma (d/2)$ is the total solid angle element, $d \widehat{\bm \sigma}$ is the solid angle element around the direction of the
unit vector $ \widehat{\bm \sigma}$, and $b_{\bm \sigma}^{-1}$ is an operator changing all the velocities ${\bm v}_{1}$ and ${\bm v}_{2}$ to its right into their precollisional values given by $b_{\bm \sigma}^{-1}{\bm v}_{1,2} = {\bm v}_{1,2}^{*} \equiv {\bm v}_{1} \mp (1+\alpha){\bm v}_{12} \cdot \widehat{\bm \sigma} \widehat{\bm \sigma} /2 \alpha$, with ${\bm v}_{12}  \equiv {\bm v}_{1}-{\bm v}_{2}$. The coefficient of normal restitution $\alpha$ is defined in the interval $0 < \alpha \leq 1$. The hydrodynamic fields are defined in terms of  $f({\bm v}_{1},t)$ in the usual way. Then, by taking velocity moments in Eq.\ (\ref{1.1}) it is seen that
\begin{equation}
\label{1.2}
\frac{\partial n}{\partial t}= \frac{\partial {\bm u}}{\partial t}=0, \quad
\frac{\partial T}{\partial t}=-\zeta (t)T,
\end{equation}
with $\zeta$ being the cooling rate given by
\begin{equation}
\label{1.3}
\zeta(t) =\frac{d+2}{4d} (1-\alpha^{2}) \nu (t).
\end{equation}
As the IBE for hard spheres and disks, Eq. (\ref{1.1}) has a similarity solution describing the homogeneous cooling state (HCS),
\begin{equation}
\label{1.4}
f_{H}({\bm v}_{1},t) =n_{H}v_{0}^{-d}(t) \chi (c_{1}),
\quad v_{0}(t) \equiv \left[ \frac{2T_{H}(t)}{m} \right]^{1/2},
\end{equation}
where $\chi(c_{1})$ is an isotropic function of the scaled velocity ${\bm c}_{1} \equiv {\bm v}_{1}/v_{0}(t)$. Substitution of Eq. (\ref{1.4}) into Eq. (\ref{1.1}), using Eq. (\ref{1.2}) yields
\begin{equation}
\label{1.5}
\frac{\widetilde{\zeta}}{2} \frac{\partial}{\partial {\bm c}_{1}} \cdot \left[ {\bm c}_{1} \chi (c_{1}) \right]
= \widetilde{J}_{M} [{\bm c}_{1} | \chi,\chi].
\end{equation}
In the above expression, $\widetilde{\zeta} \equiv \zeta_{H}(t)/\nu_{H}(t)$ and $\widetilde{J}_{M}$ is the dimensionless
version of the collision operator $J_{M}$. The index $H$ is used to characterize quantities computed in the HCS.

Next, Eq.\ (\ref{1.1}) is particularized for small (homogeneous) perturbations of the HCS. A distribution $\delta \chi$ is defined by $f({\bm v}_{1},t)= n_{H} v_{0}^{-d}(t)[\chi (c_{1})+\delta \chi ({\bm c}_{1},t)]$, with the velocities scaled relative to $v_{0}(t)$. Moreover, the time is expressed in terms of $\tau$ given by $d\tau = \nu_{H}(t) dt$. Retaining
terms up through linear order in $\delta \chi$ it is obtained
\begin{equation}
\label{1.6}
\partial_{\tau} \delta \chi ({\bm c}_{1},\tau) = \Lambda ({\bm c}_{1})\delta \chi ({\bm c}_{1},\tau),
\end{equation}
\begin{eqnarray}
\label{1.7}
\Lambda({\bm c}_{1}) \delta \chi ({\bm c}_{1},\tau) &=& \widetilde{J}_{M}[{\bm c}_{1}|\chi,\delta \chi] +
\widetilde{J}_{M} [{\bm c}_{1}|\delta \chi, \chi]- \frac{\widetilde{\zeta}}{2}
\frac{\partial}{\partial {\bm c}_{1}} \cdot \left[ {\bm c}_{1} \delta \chi ({\bm c}_{1},\tau)\right]
\nonumber \\
&& +\frac{\widetilde{\zeta}}{4}
\frac{\partial}{\partial {\bm c}_{1}} \cdot \left[ {\bm c}_{1}  \chi ( c_{1}) \right] \int d{\bm c}_{2} \left(
\frac{2 c_{2}^{2}}{d}-1 \right) \delta \chi \left({\bm c}_{2}, \tau \right).
\end{eqnarray}
Linearization of the balance equations (\ref{1.2}) around the HCS leads to
\begin{equation}
\label{1.8}
\frac{\partial \rho}{\partial  \tau}=0, \quad \frac{\partial {\bm \omega}}{\partial \tau }=\frac{\widetilde{\zeta}}{2} {\bm \omega}, \quad \frac{\partial \theta}{\partial \tau}= - \frac{\widetilde{\zeta}}{2}\, \theta - \widetilde{\zeta} \rho,
\end{equation}
with $\rho \equiv (n-n_{H})/n_{H}$, ${\bm \omega} \equiv {\bm u}/v_{0}(t)$, and $\theta \equiv (T-T_{H})/T_{H}$. The spectrum of the above equations is given by three points, $\lambda_{1}=0$, $\lambda_{2}= \widetilde{\zeta}/2$, and
$\lambda_{3} = - \widetilde{\zeta}/2$, $\lambda_{2}$ being $d$-fold degenerate. Their perturbation for finite wavevectors (gradients) in the context of the inhomogeneous IMM defines the hydrodynamic modes more generally. These eigenvalues are the same as for the IBE for hard spheres or disks \cite{BDyR03,DyB03}. In the elastic limit $\alpha \rightarrow 1$, the well known result of a $(d+2)$-fold degenerate point at zero eigenvalue is recovered.

The operator $\Lambda$ is expected to have the  corresponding hydrodynamic modes, the remaining part of the spectrum being referred to as the kinetic modes. Of course, this terminology does not preclude any mathematical difference or separation between both parts of the spectrum. Then, the eigenproblem
\begin{equation}
\label{1.9}
\Lambda ({\bm c}_{1}) \xi_{i} ({\bm c}_{1}) = \lambda_{i} \xi_{i} ({\bm c}_{1})
\end{equation}
is considered. First, attention is focussed on the hydrodynamic part of the spectrum that can be determined as follows. Define the function  $F({\bm c}_{1}, \rho, {\bm \omega},\gamma) \equiv \rho \chi (\gamma C_{1})$, with
${\bm C}_{1}={\bm c}_{1}-{\bm \omega}$. Using Eq.\ (\ref{1.5}) it follows that
\begin{equation}
\label{1.10}
\frac{\rho \widetilde{\zeta}}{2} \frac{\partial}{\partial {\bm c}_{1}} \cdot ({\bm C}_{1} F) = \gamma^{d} \widetilde{J}_{M} [{\bm c}_{1}|F,F].
\end{equation}
Differentiating this equation with respect to $\rho$, ${\bm \omega}$, and $\gamma$, taking afterwards the limit
$\rho=\gamma=1$, ${\bm \omega}=0$, it is straightforward to show that $\Lambda$ has the hydrodynamic eigenvalues
$\lambda_{i}$, $i=1,2,3$, given above, with the eigenfunctions
\begin{eqnarray}
\label{1.11}
\xi_{1}({\bm c}_{1})= (d+1) \chi({\bm c}_{1})+{\bm c}_{1} \cdot \frac{\partial \chi (c_{1})}{\partial {\bm c}_{1}}, \nonumber \\
{\bm \xi}_{2}({\bm c}_{1}) = - \frac{\partial \chi (c_{1})}{\partial {\bm c}_{1}}, \quad
\xi_{3}({\bm c}_{1}) =- \frac{\partial}{\partial {\bm c}_{1}} \cdot \left[ {\bm c}_{1} \chi (c_{1}) \right].
\end{eqnarray}
As for the hydrodynamic equations, the eigenvalue $\lambda_{2}$ is $d$-fold degenerate. The eigenfunctions $\xi_{i}$ here are the same functionals of $\chi$  as for the IBE \cite{BDyR03}. Solutions to Eq.\ (\ref{1.6}) are sought in a Hilbert space defined by the scalar product
\begin{equation}
\label{1.12}
\langle g | h \rangle \equiv \int d{\bm c}\, \chi^{-1} (c)g^{\dag} ({\bm c}) h({\bm c}),
\end{equation}
with the dagger denoting complex conjugate. The hydrodynamic eigenfunctions span a $d+2$ dimensional subspace
of the Hilbert space. They are not orthogonal, as a manifestation of $\Lambda$ being non Hermitian. Then, the
left hand eigenproblem
\begin{equation}
\label{1.13}
\Lambda^{+}({\bm c}_{1}) \overline{\xi}_{i} ({\bm c}_{1}) = \overline{\lambda}_{i} \overline{\xi}_{i} ({\bm c}_{i}),
\end{equation}
has to be considered. In this equation, $\Lambda^{+}$ is the adjoint of $\Lambda$, defined through $ \langle g| \Lambda h \rangle ^{\dag}= \langle h|\Lambda^{+} g \rangle$,
\begin{eqnarray}
\label{1.13a}
\Lambda^{+}({\bm c}_{1}) g({\bm c}_{1}) &=& \frac{d+2}{2 \Omega_{d}} \int d{\bm c}_{2} \int d {\bm \sigma}\,
\chi (c_{1}) \chi (c_{2}) \left( b_{\bm \sigma} -1 \right) \left[ \chi^{-1}(c_{1}) g({\bm c}_{1})+ \chi^{-1}(c_{2}) g({\bm c}_{2}) \right] \nonumber \\
&& + \frac{\widetilde{\zeta}}{4}\, \chi (c_{1})  \left( \frac{2 c_{1}^{2}}{d}-1 \right) \int d{\bm c}_{2} \chi^{-1} (c_{2})
g({\bm c}_{2}) \frac{\partial}{\partial {\bm c}_{2}} \left[ {\bm c}_{2} \chi (c_{2}) \right] \nonumber \\
&& + \frac{\widetilde{\zeta}}{2}\, \chi (c_{1}) {\bm c}_{1} \cdot \frac{\partial}{\partial {\bm c}_{1}} \left[ \chi^{-1}(c_{1}) g({\bm c}_{1}) \right],
\end{eqnarray}
$b_{\bm \sigma}$ being the operator inverse of $b_{\bm \sigma}^{-1}$. Although there seems to be no simple relationship between the functions $\xi_{i}$ and $\overline{\xi}_{i}$, the hydrodynamic part of the spectrum of $\Lambda^{+}$ has been found by simple inspection. The functions
\begin{equation}
\label{1.14}
\overline{\xi}_{1} ({\bm c}_{1}) = \chi (c_{1}), \quad \overline{\bm \xi}_{2} ({\bm c}_{1}) ={\bm c}_{1} \chi ( c_{1}), \quad \overline{\xi}_{3}({\bm c}_{1}) =\left( \frac{c_{1}^{2}}{d}+\frac{1}{2} \right) \chi (c_{1}),
\end{equation}
are solutions of Eq. (\ref{1.13}), corresponding to the eigenvalues $\overline{\lambda}_{i} = \lambda_{i}$, $i=1,2,3$, respectively.
Again, the eigenvalue $\widetilde{\zeta}/2$ is $d$-fold degenerate. Moreover, it is easily verified that
\begin{equation}
\label{1.15}
\langle \overline{\xi}_{i} | \xi_{j} \rangle = \delta_{ij},
\end{equation}
for the hydrodynamic eigenfunctions given by Eqs.\  (\ref{1.11}) and (\ref{1.14}).

It is also possible to get some information about the remaining, kinetic part of the spectrum of $\Lambda^{+}$. Here attention will be restricted to two eigenfunctions which have a clear physical interpretation. A direct evaluation gives
\begin{equation}
\label{1.16}
\Lambda^{+} ({\bm c}_{1}) c_{1x}c_{1y} \chi (c_{1}) = - \frac{1}{4} (1+ \alpha)^{2} c_{1x}c_{1y} \chi (c_{1}).
\end{equation}
It follows that $\overline{\xi}_{4} ({\bm c}_{1}) = c_{1x}c_{1y} \chi (c_{1})$ is an eigenfunction of $\Lambda^{+}$
with eigenvalue $\overline{\lambda}_{4} = -(1+\alpha)^{2}  /4$. Note that $\overline{\xi}_{4}$ is proportional to the dynamical variable whose velocity average provides the pressure tensor \cite{Mc89}. It is important to check whether
the kinetic eigenvalue $\overline{\lambda}_{4}$ is actually separated from the hydrodynamic part of the spectrum, as required for the validity of the hydrodynamic description. Note that if it is assumed that any function of the Hilbert space can be expanded in terms of the eigenfunctions of $\Lambda$, then the left eigenvalues are also right eigenvalues, i.e. there is a solution of Eq.\ (\ref{1.8}) with $\lambda_{4}= \overline{\lambda}_{4}$. Consequently, the general solution of Eq. (\ref{1.6}) contains a contribution proportional to $\exp (\overline{\lambda}_{4} \tau) \xi_{4}({\bm c}_{1})$, where $\xi_{4}({\bm c}_{1})$ is the (unknown) eigenfunction  of $\Lambda$ with eigenvalue $\lambda_{4}= \overline{\lambda}_{4}$, or a linear combination of eigenfunctions in the case of degeneracy. The validity of hydrodynamics requires that the above contribution decay faster than all the hydrodynamic modes, i.e. it must be $| \overline{\lambda}_{4}| >
| \lambda_{3}| = \widetilde{\zeta}/2$. Taking into account the exact result given by Eq.\ (\ref{1.4}), this condition is equivalent to $\alpha > \alpha_{1}$ with $\alpha_{1}=-(d-2)/(3d+2)$. For $d=3$ this expression leads to an
unphysical negative value of $\alpha_{1}$, while for $d=2$ it is $\alpha_{1} =0$.

\begin{figure}
\includegraphics[scale=0.7,angle=0]{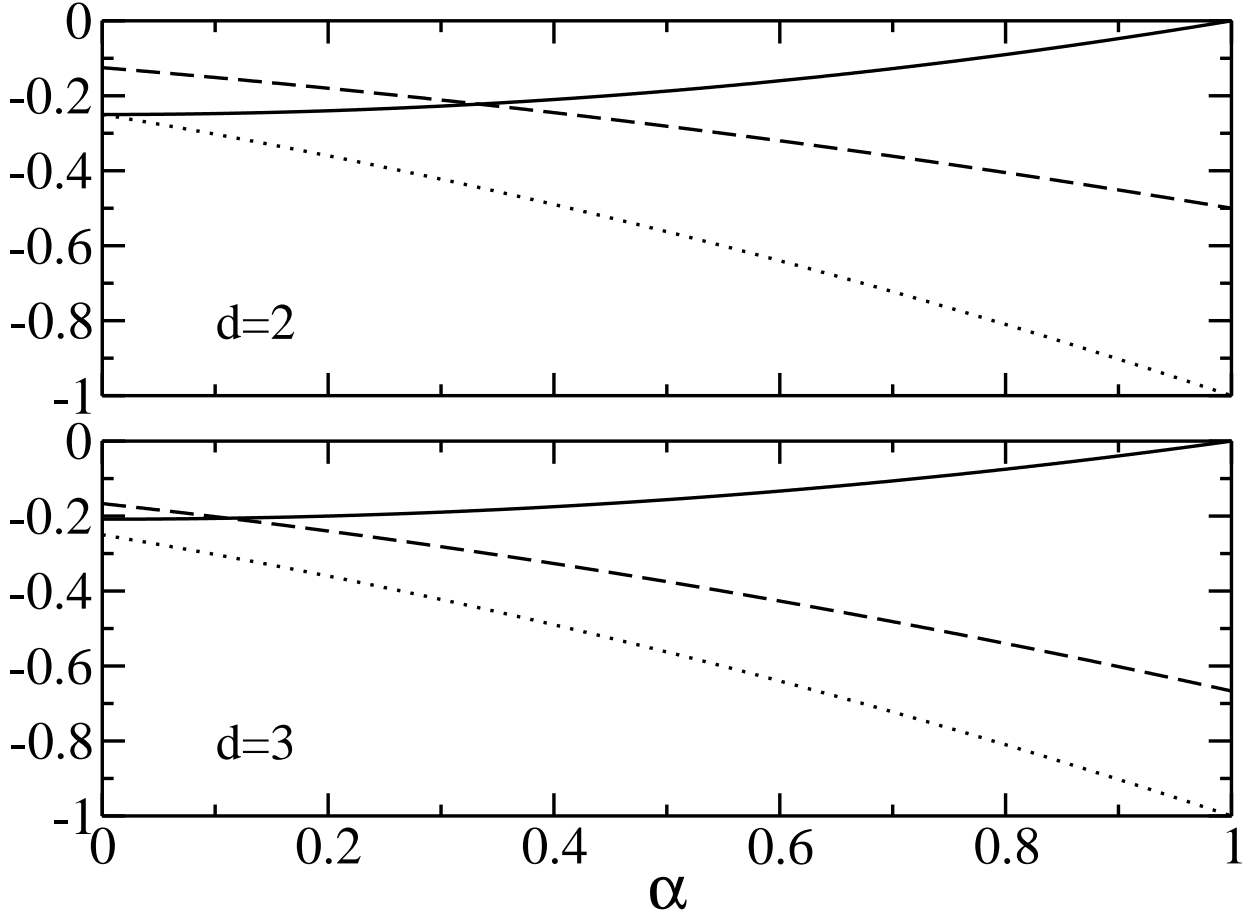}
\caption{The two kinetic eigenvalues $\overline{\lambda}_{4}$ (dotted lines) and $\overline{\lambda}_{5}$ (dashed lines), and the slowest hydrodynamic mode $\lambda_{3}$ (solid lines) versus the
coefficient of normal restitution $\alpha$ for a system of inelastic Maxwell molecules. Note that  the three modes are negative. } \label{fig1}
\end{figure}

The above result suggests to check whether there is another left eigenfunction proportional to the dynamical variable
providing the heat flux. One reason to consider these fluxes as candidates to be eigenfunctions of $\Lambda^{+}$ is that they are known to be orthogonal to the
hydrodynamic modes $\xi_{i}$ \cite{BDyR03}. Explicit evaluation gives that
\begin{equation}
\label{1.17}
\overline{\xi}_{5}({\bm c}_{1}) = \left( c_{1}^{2}- \frac{d+2}{2} \right) c_{1x} \chi(c_{1})
\end{equation}
is a solution of Eq.\ (\ref{1.13}) with
\begin{equation}
\label{1.18}
\overline{\lambda}_{5}= - \frac{(d-1)(1+\alpha)^{2}}{4d}.
\end{equation}
The condition for the validity of hydrodynamics following from the existence of this kinetic mode is that
$\alpha > \alpha_{2} = (4-d)/3d$. Therefore, there is a finite value of the coefficient of normal restitution
below which there is no time scale separation between the hydrodynamic and the kinetic parts of the distribution function. For $d=2$ it is $\alpha_{2}=1/3$ and for $d=3$ $\alpha_{2}= 1/9$. In Fig. \ref{fig1} the two kinetic eigenvalues
$\overline{\lambda}_{4}$ and $\overline{\lambda}_{5}$ are compared with the slowest hydrodynamic mode $\lambda_{3}$ for $d=3$.

The lack of dominance of the hydrodynamic modes over the mode related with the heat flux for long times, reflects itself in the calculation of the Navier-Stokes heat conductivity of an inelastic  gas of Maxwell molecules. As already mentioned, the form of the hydrodynamic spectrum for IMM reported here is the same as for the IBE for hard spheres. The only difference is in the explicit expression of the distribution function of the HCS and in
the value of the cooling rate. As a consequence, the formal expressions for the transport coefficients derived in \cite{BDyR03,DyB02} from the IBE, can be directly applied to the case of Maxwell molecules. Then, the time-dependent
heat conductivity $\kappa$ is given by
\begin{equation}
\label{1.19}
\kappa (\tau) = n_{H}  \frac{ v_{0}^{2}(t)}{\nu (t)} \int_{0}^{\tau} d \tau^{\prime}\, \widetilde{\kappa} (\tau^{\prime}),
\end{equation}
\begin{eqnarray}
\label{1.20}
\widetilde{\kappa}(\tau)  & = & \frac{1}{2} \int d{\bm c}_{1}\ \left( c_{1}^{2}-\frac{d+2}{2} \right) c_{1x} e^{\tau \left( \Lambda + \widetilde{\zeta}/2 \right)} \xi_{3} ({\bm c}_{1}) c_{1x}
\nonumber \\
&=& \frac{e^{\tau \left( \lambda_{5}+\widetilde{\zeta}/2 \right)}}{2} \int d{\bm c}_{1}\, \xi_{3} ({\bm c}_{1}) \left( c_{1}^{2} - \frac{d+2}{2} \right) c_{1x}^{2}.
\end{eqnarray}
The existence of hydrodynamics to Navier-Stokes order requires that the correlation function $\widetilde{\kappa} (\tau)$ decay to zero for $\tau \gg 1$. This clearly implies that $|\lambda_{5}| > \widetilde{\zeta} /2$, i.e.
$\alpha > \alpha_{2}$. Consistently with this analysis, it is found that the heat conductivity for inelastic Maxwell molecules obtained by the Chapman-Enskog procedure also diverges for $\alpha \rightarrow \alpha_{2}$ \cite{Sa03}.

To put the results reported here in a proper context some comments seem appropriate. (i) It has been proven that the solution of the Boltzmann equation for IMM tends to the HCS for arbitrary initial conditions \cite{BCyT02}. The
lack of time scale separation discussed here does not contradict this general property, since the obtained kinetic eigenvalues are negative, therefore leading to decaying in time contributions. (ii) It is worth to stress that
the failure of hydrodynamics in IMM should not be understood as limited to the Navier-Stokes approximation, but to
any closed description of the system in terms of the hydrodynamic fields. (iii) Of course, the value of $\alpha$ for which the scale separation actually fails is not known, but only a lower bound has been determined. It seems sensible to expect higher limiting values of $\alpha$ associated to eigenfunctions involving higher velocity powers.
(iv) On the other hand, there is no reason to expect a similar behavior for the IBE for hard spheres or disks. On the contrary, there are some indications that
this is not the case. The expressions derived for the Navier-Stokes transport coefficients are regular functions of $\alpha$ for $0 < \alpha \leq 1$ \cite{BRMyG05}, and the two kinetic modes considered here are not left eigenfunctions of the linearized inelastic Boltzmann operator. Actually, other relevant deep differences between the IBE and the IMM have
been found, as for instance the asymptotic decay of the distribution function of the HCS for large velocities \cite{EyB02,EyP97}.

This research was partially supported by the Ministerio de
Educaci\'{o}n y Ciencia (Spain) through Grant No. FIS2008-01339 (partially financed
by FEDER funds).

\end{document}